\documentclass[a4paper]{article}

\usepackage{atmohead2013}
\usepackage[english]{babel}

\title{The Atmospheric Monitoring system of the JEM-EUSO telescope}

\shorttitle{The Atmospheric Monitoring system of the JEM-EUSO telescope}

\authors{
S. Toscano$^{1}$,
A. Neronov$^{1}$,
M.D. Rodr\'iguez Fr\'ias$^{2}$,
S. Wada$^{3}$
for the JEM-EUSO Collaboration.
}

\afiliations{
$^1$ ISDC Data Centre for Astrophysics, Versoix, Switzerland \\
$^2$ SPace \& AStroparticle (SPAS) Group, UAH, Madrid, Spain \\
$^3$ RIKEN Advanced Science Institute, Japan\\
}

\email{Simona.Toscano@unige.ch}

\abstract{The JEM-EUSO observatory on board of the International Space Station (ISS) is a proposed pioneering space mission devoted to the investigation of Ultra High Energy Cosmic Rays (UHECRs). Looking downward at the earth's atmosphere with a 60$^\circ$ Field of View (FoV), the JEM-EUSO telescope will detect the fluorescence and Cherenkov UV emission from UHECR induced Extensive Air Showers (EAS) penetrating in the atmosphere.ÊThe capability of reconstructing the properties of the primary cosmic ray depends on the accurate measurement of the atmospheric conditions in the region of EAS development. The Atmospheric Monitoring system of JEM-EUSO will continuously monitor the atmosphere at the location of the EAS candidates and between the EAS and the JEM-EUSO telescope. With an UV LIDAR and an Infrared (IR) Camera the system will monitor the cloud cover and retrieve the cloud top altitude with an accuracy of $\sim$ 500 m and the optical depth profile of the atmosphere with an accuracy of $\Delta\tau \leq$ 0.15 and a resolution of 500 m.Ê
In this contribution the Atmospheric Monitoring system of JEM-EUSO will be presented. After a brief description of the system, the capability to recover the cloud top height and optical depth and to reconstruct the shower profile will be shown based on satellites data and simulation studies.}


\begin{document}
\maketitle

\section{Introduction}
\begin{figure}[h]
  \centering
  \includegraphics[width=0.4\textwidth]{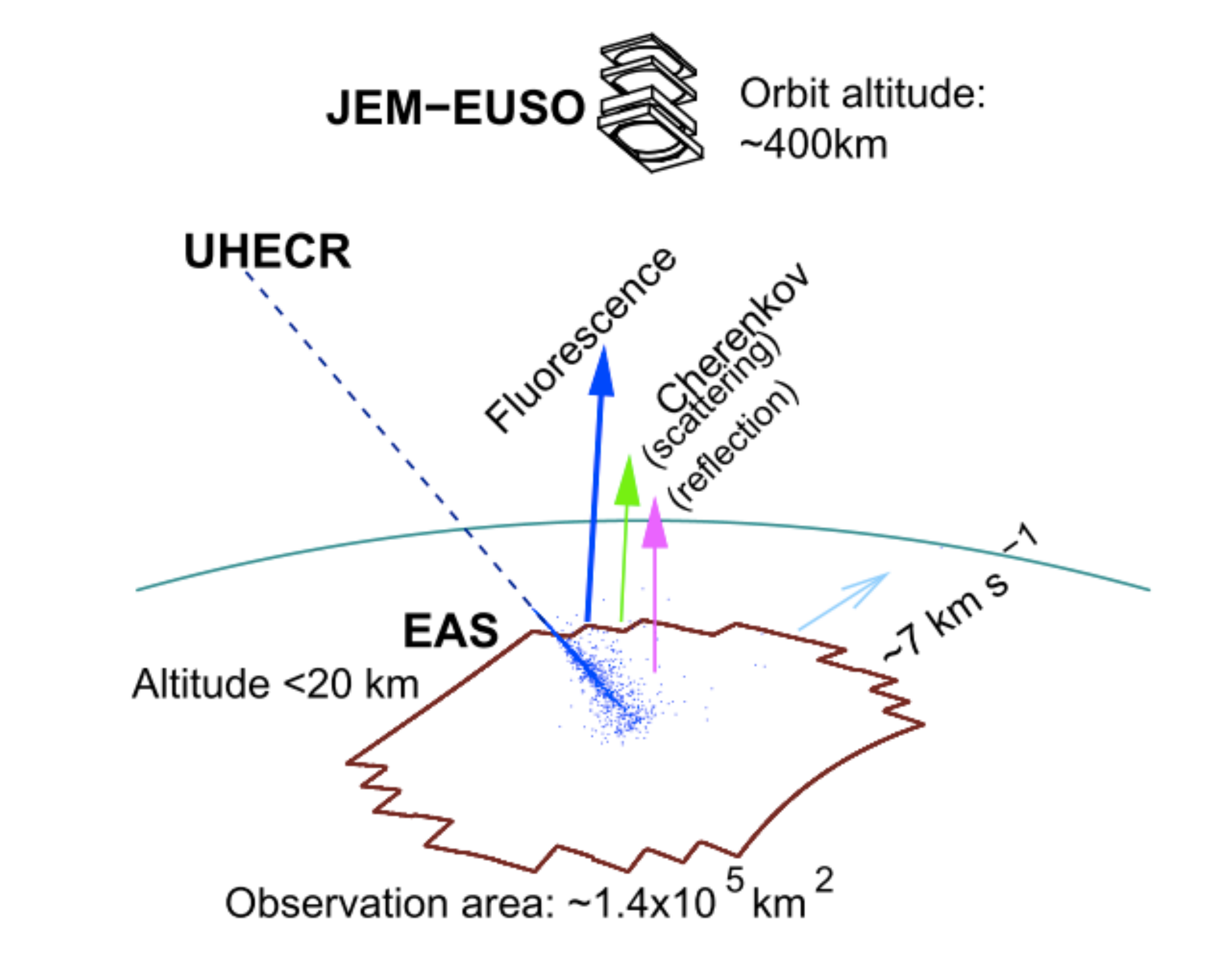}
  \caption{Illustration of UHECR observation principle in the JEM-EUSO mission (taken from \cite{ExposurePaper}).}
  \label{fig:detection_principle}
 \end{figure}

The Extreme Universe Space Observatory on board of the Japanese Experiment Module (JEM-EUSO) is a new space-based mission to study the origin and nature of Ultra High Energy Cosmic Rays (UHECRs) \cite{Picozza_ICRC2013}. JEM-EUSO will orbit at an altitude of $\sim$ 400 km detecting the faint UV fluorescence and Cherenkov light emitted by Extensive Air Showers (EAS) initiated by primary UHECRs hitting the earth's atmosphere. The detection principle of the JEM-EUSO telescope is shown in Fig.~\ref{fig:detection_principle}.
 
\noindent
Secondary charged particles, predominantly electrons, excite atmospheric nitrogen molecules along the EAS track. The UV fluorescence light, emitted isotropically as a result of the de-excitation process, represent the dominant component of the detected signal. The Cherenkov component represents a small correction for the overall energy detected by the telescope. Since the Cherenkov light is forward-beamed, photons are observed only by the scattering in the atmosphere or diffuse reflection from the ground or a cloud-top (known as ``ground mark" and ``cloud mark"). The timing of the Cherenkov mark provides a useful constrain to reconstruct the geometry of the track and to identify the type of the primary particle that generated the EAS. \\
The amount of fluorescence and Cherenkov light reaching JEM-EUSO depends on the absorption and scattering properties of the atmosphere. A correct reconstruction of the UHECR energy and of the type of the primary cosmic ray particle requires, therefore, information about absorption and scattering of the UV light. Moreover, the presence of clouds and aerosol layers distorts the UV signal from the EAS leading to systematic errors in the reconstruction of the shower profile. Errors in the reconstruction of the EAS profile can finally affect the detector aperture and duty cycle.\\
Moving with the ISS at a velocity of $\sim7$ km/sec, the detector ``footprint" will cross all possible weather conditions. The Atmospheric monitoring system of JEM-EUSO will continuously monitor the variable atmospheric conditions inside the JEM-EUSO FoV, using a LIDAR and an Infrared camera. \\
 
\section{The Atmospheric Monitoring system}

The Atmospheric Monitoring (AM) system of JEM-EUSO will include: 
\begin{enumerate}
\item an {\bf Infrared (IR) camera} to get information on the cloud cover and the measurement of the top altitude, at least for the optically thick clouds; 
\item a {\bf LIght Detection And Ranging (LIDAR)} device which will be used for the detection of aerosol layers and clouds (especially optically thin) and the measurement of cloud altitude and optical depth;
\item {\bf global atmospheric models} generated from the analysis of all available meteorological data by forecasting services such as the National Centers for Environmental Predictions (NCEP) \cite{NCEP}, the Global Modeling and Assimilation Office (GMAO) \cite{GMAO} and the European Centre for Medium-Range Weather Forecasts (ECMWF) \cite{ECMWF};
\item the {\bf slow mode data} of JEM-EUSO, the monitoring of the pixel signal rate every 3.5 sec for the observation of transient luminous events (TLEs), which will give additional information on cloud distribution and the intensity of the night sky airglow.
\item the {\bf Global Light System} (GLS), a network of calibrated UV light sources on ground, as a part of the JEM-EUSO calibration system, which will be used to monitor and validate key parameters of the detector \cite{GLS}. The GLS will provide additional information on atmospheric conditions of some specific ground based sites, as well as an independent verification of some of the cloud free conditions in the JEM-EUSO FoV.
\end{enumerate}  

\begin{figure}[h]
  \centering
  \includegraphics[width=0.5\textwidth]{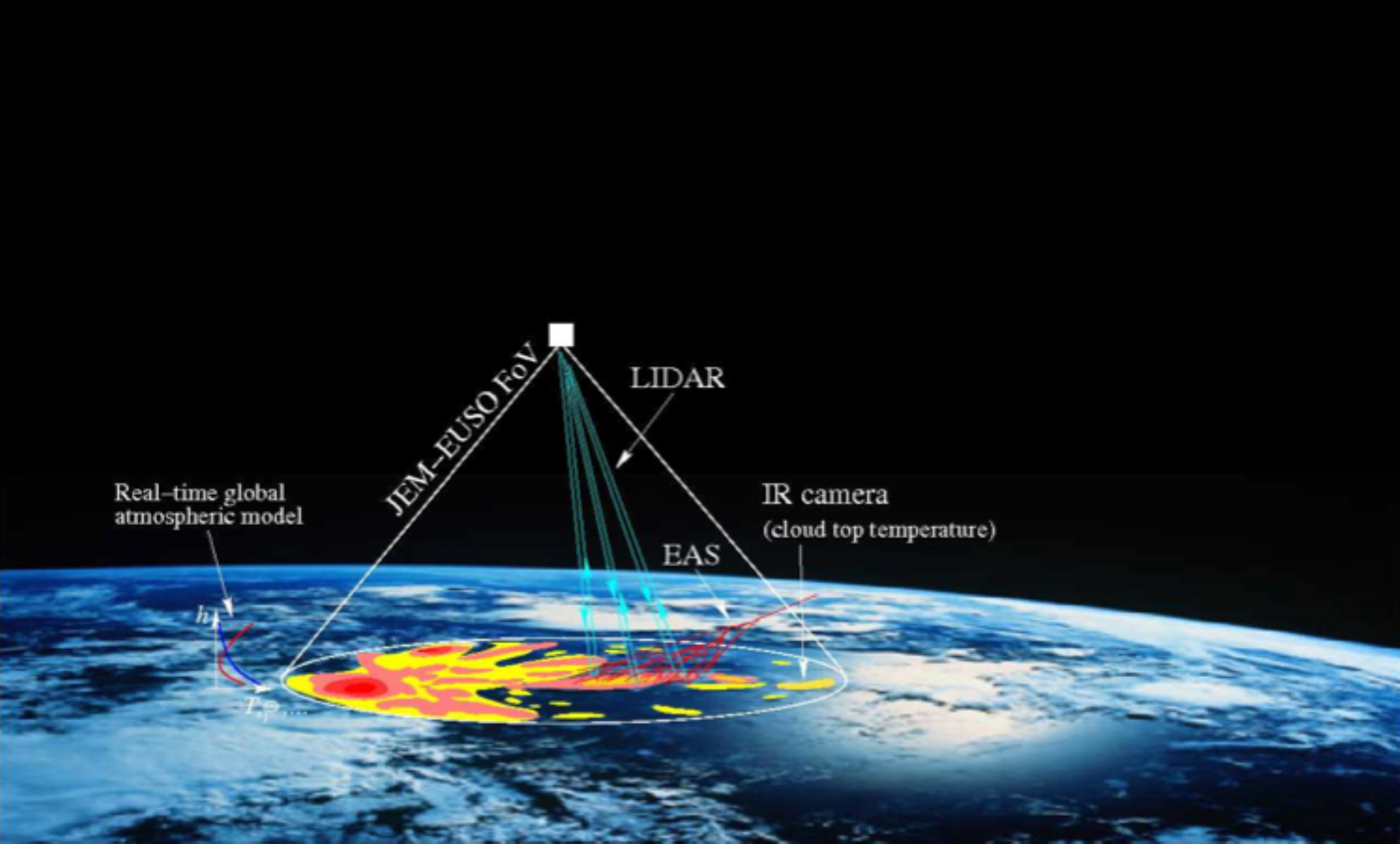}
  \caption{Sketch of the concept of Atmospheric Monitoring in JEM-EUSO.}
  \label{fig:AMS}
 \end{figure}

The detection principle of the AM system is illustrated in Fig.~\ref{fig:AMS}. 
The JEM-EUSO telescope will observe the EAS development only during night time. The IR camera will cover the entire FoV of the telescope to detect the presence of clouds and to obtain the cloud cover and the cloud top altitude. The LIDAR will shoot in some pre-defined position around the location of triggered EAS candidates and at GLS stations. The LIDAR will be used to measure the clouds altitude and optical depth as well as the optical depth vertical profile of the atmosphere along these directions with a range accuracy of 375 m in nadir. The IR camera and the LIDAR have been designed to work in a complementary way. On one hand the IR camera will give information on the cloud cover over the entire FoV, which is not accessible with the LIDAR, since this device can measure the optical properties of the atmosphere only in a given direction. On the other hand, the LIDAR will     
measure the optical properties of optically thin clouds which are difficult to identify with the IR camera. The analysis of the atmospheric data will be complemented by the use of the atmospheric global models and the slow mode data of JEM-EUSO. 

Statistical studies show that $\sim$ 70\% of the EAS triggered events will develop in cloudy conditions \cite{Mario_ICRC2011, Lupe2013}, meaning that only $\sim$ 30\% of the triggered events will be tagged as ``golden" and used for physics analysis if no correction is applied to the cloud-affected events. The minimal task of the AM system will be to provide such a criteria for the final selection of the ``golden sample" \footnote{Triggered events developing in clear sky conditions}; while a more advanced task will be to improve the reconstruction of those cloud-affected shower profiles which could be retained for further analysis. In order to achieve these goals the requirements on the precision of the atmospheric measurements have been set\footnote{The reported numbers are derived from the general requirements on the precision of the UHECR measurement as explained in \cite{Andrii_ICRC2013}} such that: 
\begin{itemize}
\item the transmission uncertainty has to be smaller than $\sim 15$~\% at nadir;
\item the measurement in the accuracy of the cloud top altitude has to be $\Delta H \leq$~500~m.
\end{itemize}

\section{The Infrared Camera}
\begin{figure}[h]
  \centering
  \includegraphics[width=0.5\textwidth]{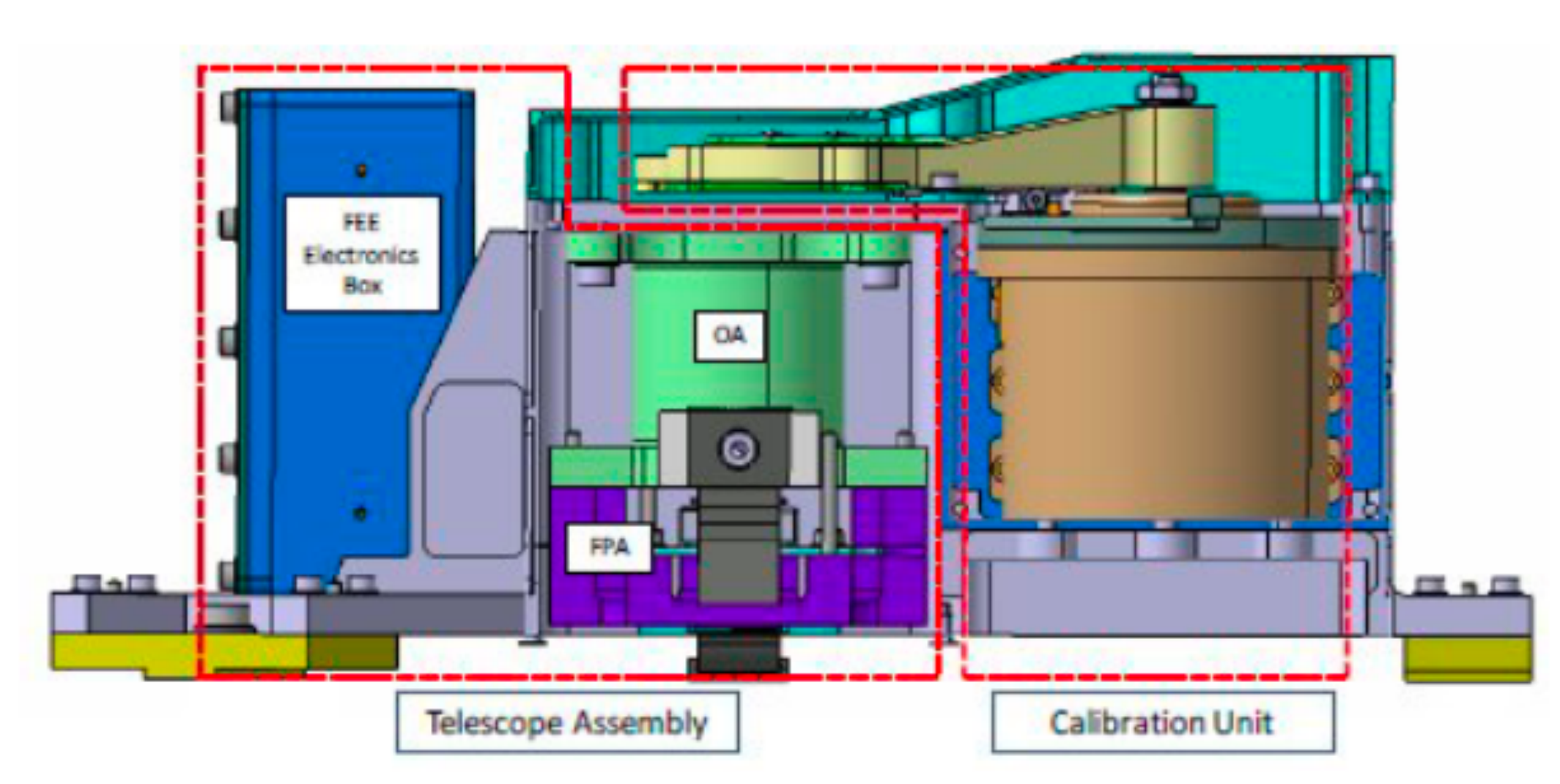}
  \caption{Illustration of the preliminary design of the IR camera telescope assembly. }
  \label{fig:IRCamera}
 \end{figure}

The IR camera of the AM monitoring system of JEM-EUSO  is a micro-bolometer based infrared imaging system designed to measure the cloud cover and the cloud top altitude during the observation period of the JEM-EUSO main instrument. The measurement of the temperature of the cloud will be used to estimate the altitude of the cloud top layers. Such a measurement is possible in troposphere (0-10 km) where the temperature gradient is stable and equal to $dT/dH \simeq 6^\circ$/km. To achieve the precision of measurement of the cloud top altitude imposed by the requirements, the precision of the temperature measurement of the IR camera has to be: $\Delta T = (dT / dH) \Delta H = 3 K$. Additional information, such as the temperature profiles, will be obtained from the GDAS model \cite{GDAS}, which provides profiles of several state variables on a $1^\circ \times1^\circ$ grid several times a day by extrapolating from thousands of daily radiosonde measurements from around the globe. \\
The preliminary design of the camera is described in \cite{IRCamera_ICRC2013}; it comprises three main blocks: 
\begin{itemize}
\item the Telescope Assembly, illustrated in Fig.~\ref{fig:IRCamera} acquires the infrared radiation using an uncooled micro-bolometer and it converts the radiation into digital counts;
\item the Electronic Assembly processes and transmits the images and it provides the electrical system and the thermal control;
\item the Calibration Unit is used to ensure the high absolute temperature accuracy ($\Delta T \leq$ 3 K) needed to fulfill the requirements ($\Delta H \leq $~500~m).  
\end{itemize}

 \noindent
To provide the cloud cover and to retrieve the Cloud Top Height (CTH) different algorithms, based on the radiance measured by the IR camera, have been developed within the collaboration. The CTH can be retrieved either using stereo vision techniques or accurate radiometric information \cite{Anna_ICRC2013}. 
 
 \subsection{Stereo vision algorithm}

\begin{figure}[h]
  \centering
  \includegraphics[width=0.4\textwidth]{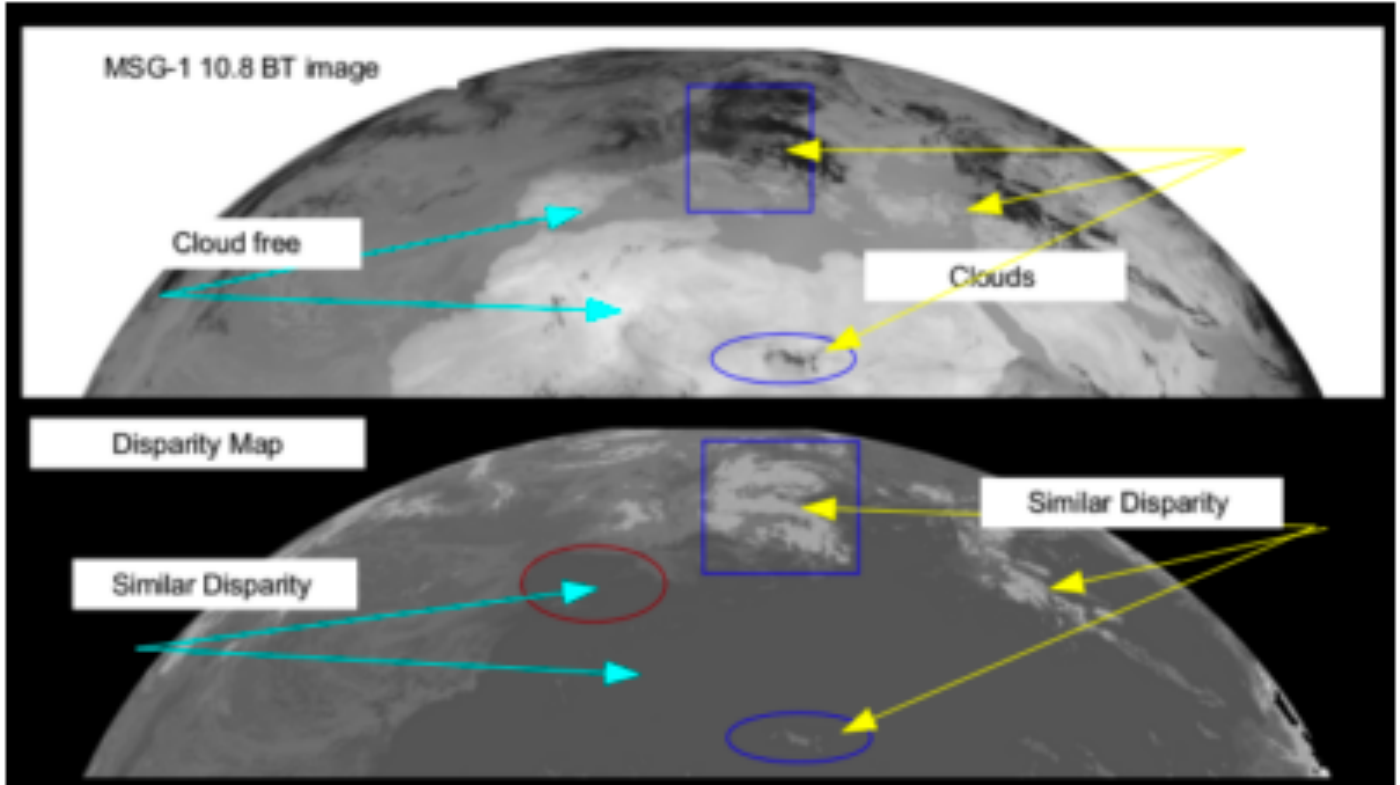}
  \caption{Disparity map from \cite{Anna_ICRC2013}. \emph{Top:} MSG-8 image used to estimate the map; \emph{Bottom:} Corresponding disparity map. The brightest points represent the regions closer to the camera sensor and therefore the highest from the ground.}
  \label{fig:Stereo}
 \end{figure}

The stereo imaging is accomplished by one camera moving along the observed scene, exploiting the ISS displacement. The scene results imaged from two different views and the intersection is processed to retrieve the distance from the IR device (depth). The stereo images are acquired in two different bands.
Fig.~\ref{fig:Stereo} shows an example of disparity (parallax effect) map for a ``mono band stereo system" composed by the Meteosat Second Generation geostationary satellites MSG-8 and MSG-9. Areas having the same grey level represent points with the same distance from the sensor of the camera; the more distant is the object the smallest is the disparity.

 \subsection{Temperature retrieval algorithms} 
\begin{figure}[h]
  \centering
  \includegraphics[width=0.5\textwidth]{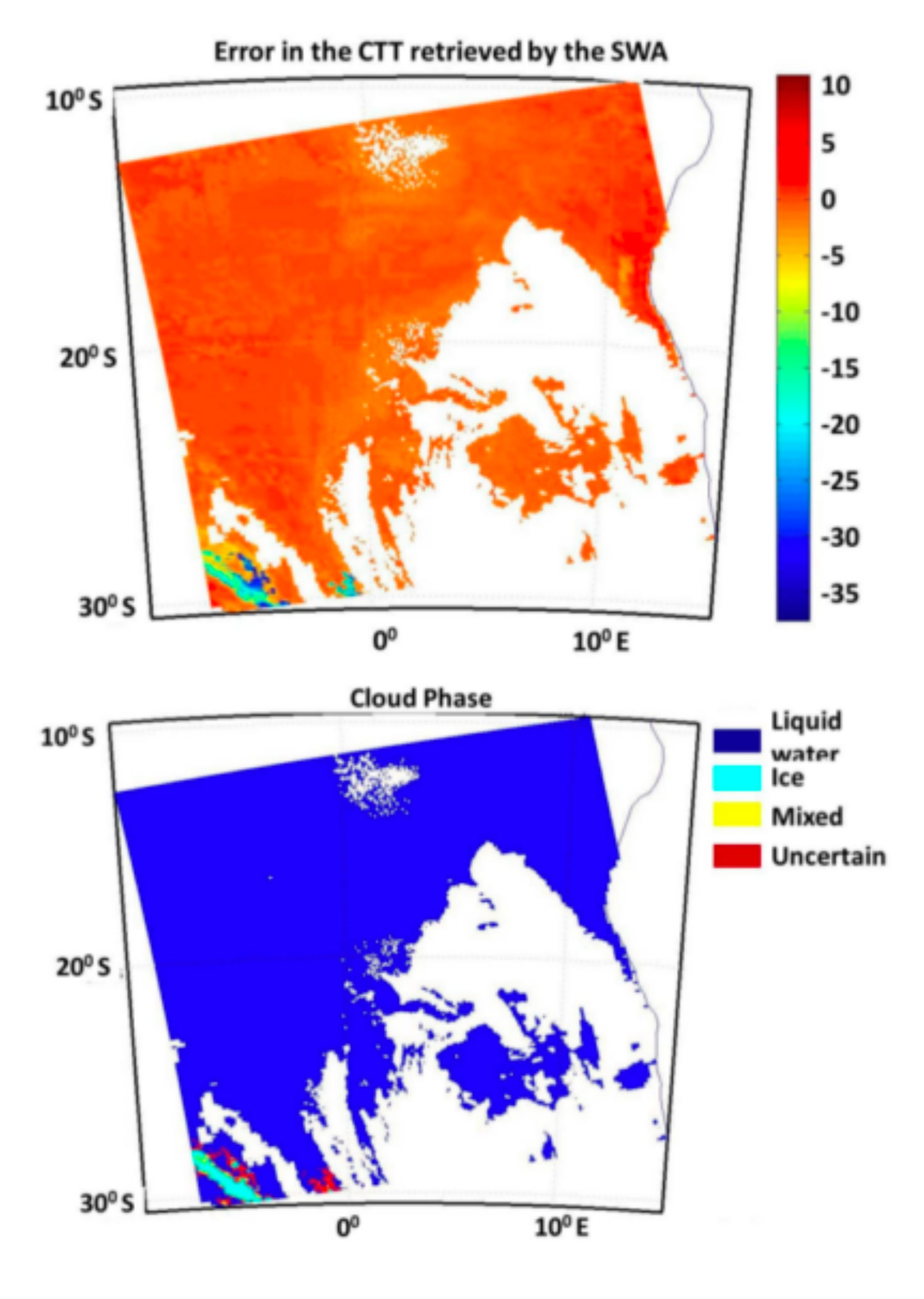}
  \caption{Temperature retrieval errors (top) from the SWA and cloud phase (bottom) provided by MODIS (from \cite{Anna_ICRC2013}).}
  \label{fig:SWA}
 \end{figure}

Due to absorption and emission processes in the atmosphere, the radiance measured by the IR camera is not the radiance emitted by the clouds. Specific algorithms are needed to correct for these atmospheric effects and to retrieve the Brightness Temperature (BT) of the cloud from the measured radiance.  \\
A Split Window Algorithm (SWA) \cite{Inoue, Nauss} has been successfully applied to retrieve the cloud top temperature of water clouds from the BT as it will be measured by the IR camera \cite{Anna_ICRC2013}. Fig.~\ref{fig:SWA} shows the temperature retrieval errors obtained when the SWA is applied to a MODIS \cite{MODIS} image. The correlation of the errors with the cloud phase provided by MODIS shows that the SWA is able to retrieve with high accuracy ($\sim$~1~K) the temperature of thick water clouds. Additional algorithms are needed for the remaining cases (thin clouds, ice clouds, broken clouds) and they are currently under study.  

\section{The LIDAR}
\begin{figure}[h]
  \centering
  \includegraphics[width=0.5\textwidth]{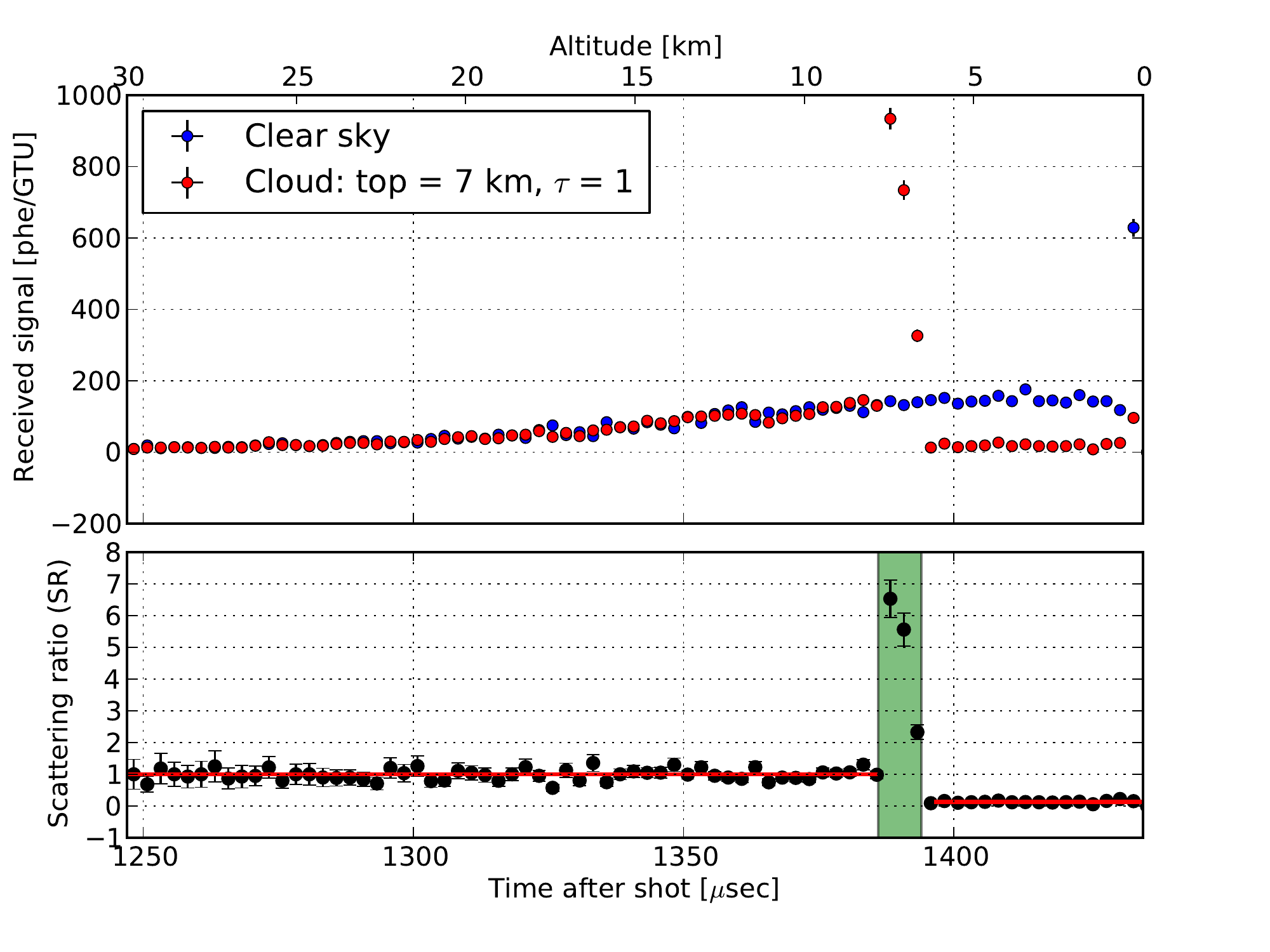}
  \caption{\emph{Top}: LIDAR backscattered signal in clear sky (blue) and in the presence of an optically thick ($\tau = 1$) cloud (red) as a function of the time after shooting the laser and the altitude. \emph{Bottom}: Scattering ratio (SR).}
  \label{fig:LIDAR}
 \end{figure}

\begin{figure}[h]
  \centering
  \includegraphics[width=0.5\textwidth]{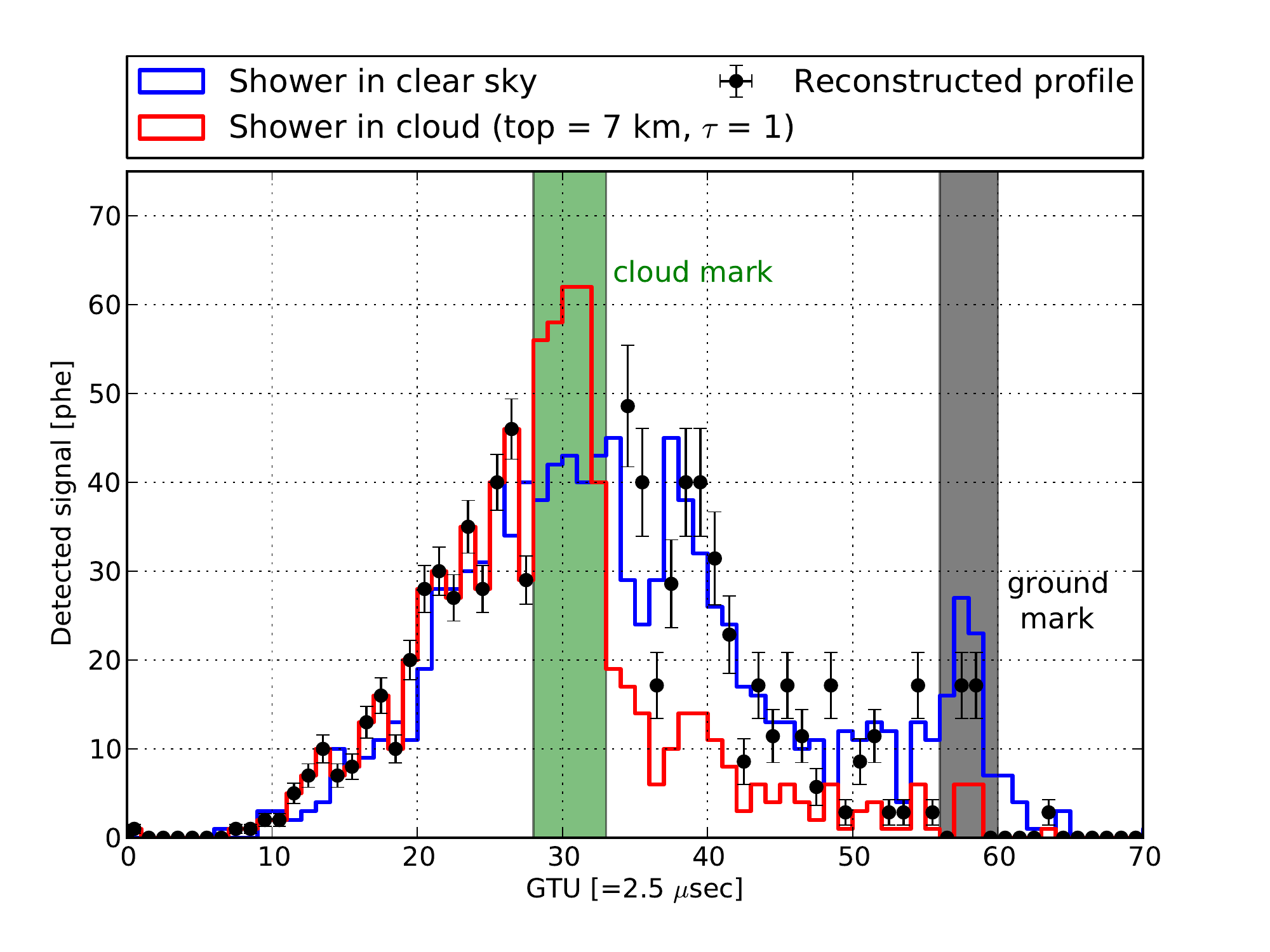}
  \caption{Reconstructed time profile (black points) of $10^{20}$ eV EAS together with the clear sky (blue) and cloud affected (red) profiles. Error bars are statistical only.}
  \label{fig:RecoProfile}
 \end{figure}

The LIDAR is composed of a transmission and receiving system. The transmission system comprises a Nd:YAG laser and a pointing mechanism to steer the laser beam in the direction of triggered EAS events. The operational wavelength is the third harmonic of the Nd:YAG laser, at $\lambda = $~355~nm. The pointing mechanism has been designed to have a steering mirror with two angular degrees of freedom and a maximal tilting angle of $\pm$~15$^\circ$, needed to move the laser beam anywhere within the JEM-EUSO FoV. The laser backscattered signal will be received by the main JEM-EUSO telescope, which is suited for the detection of the 355 nm wavelength. A summary of the specifications needed for the entire system can be found in \cite{Andrii_ICRC2013}.  \\
Simulation \cite{Toscano_ICRC2013} have been carried out to study the capability of the system in retrieving the physical properties of atmospheric features such as clouds or aerosol layers. Fig.~\ref{fig:LIDAR} shows an example of the simulated laser backscattered signal as it would appear in the JEM-EUSO detector. 
The top panel shows the comparison of the signal in case of clear sky (blue points) and in presence of a cloud (red points) as a function of the time after shooting the laser and the altitude. In this example the LIDAR is able to detect the presence of a cloud at $\sim$~7~ km, assumed horizontally uniform, as an increasing of backscattered signal coming from that region. The bottom panel shows the so-called LIDAR Scattering Ratio (SR), the ratio between the backscattered signal detected in real conditions and a reference profile represented by the backscattered signal in clear sky. 
Fitting the SR in the region below the cloud allows for the measurement of the optical depth simply using the formula $\tau = -log(SR) / 2$. 
\noindent 
Using the LIDAR measurement of the optical depth it is possible to retrieve the EAS profile distorted by the presence of the cloud. Once the cloud is detected by the instrument it is possible to correct the shower to reproduce the profile how it would appear in a purely molecular atmosphere. Fig.~\ref{fig:RecoProfile} shows the photons arrival time in GTU\footnote{The Gate Time Unit, or GTU, is the time unit of the detector focal surface; 1~GTU corresponds to 2.5~$\mu$sec.} at the detector focal plane for a shower generated by a UHE proton with $E = 10^{20}$~eV and $\theta = 60^\circ$. The blue histogram represent the profile of the shower developing in clear sky conditions, characterized by the presence of the ``ground mark" at $\sim$60~GTU. This feature is due to Cherenkov photons hitting the ground and reflected back to the JEM-EUSO focal surface. The red histogram shows the profile of the shower crossing the same optically thick cloud shot by the LIDAR and located at an altitude of $\sim$7~km. As in the case of the LIDAR, the presence of the cloud modify the EAS time profile, with the appearance of a new feature (the ``cloud mark") at $\sim$~28~GTU, and the ground mark vanishing. After the correction is done using the LIDAR measurement it is possible to retrieve the correct profile (black points) and almost entirely recover the ground mark feature.     

\section{Conclusions}
The JEM-EUSO telescope will monitor the earth's atmosphere from space looking for EAS generated by the passage of UHECRs hitting the Earth. The detector footprint will experience all possible weather conditions during the overall data taking period. The Atmospheric Monitoring system designed for the JEM-EUSO experiment will continuously monitor the telescope FoV in order to get the cloud cover and cloud top height as well as the measurement of the optical depth profile of the atmosphere around the EAS location. The information coming from the IR camera and the LIDAR will be complemented by the use of Global atmospheric models and the JEM-EUSO slow data. The atmospheric data will be used to correct the profiles of cloud-affected EAS events for the effect of clouds and aerosol layers, so that most of them could be retained for the UHECR data analysis. \\  
The results presented is this paper take into account only statistical uncertainties. Systematic errors due to the inversion method, the use of global models and of a not comprehensive simulation of the multiple scattering process in the atmosphere will increase the global uncertainties. To be able to retain the highest number of UHECR events for the final data analysis an estimation of the systematic uncertainties is needed.

\end{document}